\documentclass[%
 reprint,
superscriptaddress,
 amsmath,amssymb,
 aps,
prl,
]{revtex4-2}

\usepackage{graphicx}% Include figure files
\usepackage{dcolumn}% Align table columns on decimal point
\usepackage{bm}% bold math
\usepackage[normalem]{ulem}
\usepackage{xcolor}
\usepackage{multirow}
\usepackage{booktabs}
\usepackage{titlesec}
\titleformat{\section}[runin]{\normalfont}{\thesection.}{0em}{}[]
\titlespacing*{\section}
{0pt}{\baselineskip}{0em}
\newcommand{\mysection}[1]{\section*{\indent\textit{#1}.---}}

\begin{document}

\title{Ringdown gravitational waves from close scattering of two black holes}

\author{Yeong-Bok Bae}
\thanks{These authors contributed equally to this work: \\ astrobyb@gmail.com, younghwan.hyun@gmail.com}
\affiliation{Particle Theory and Cosmology Group, Center for Theoretical Physics of the Universe, Institute for Basic Science (IBS), Daejeon, 34126, Republic of Korea}

\author{Young-Hwan Hyun}
\thanks{These authors contributed equally to this work: \\ astrobyb@gmail.com, younghwan.hyun@gmail.com}
\affiliation{Korea Astronomy and Space Science Institute (KASI), 776 Daedeok-daero, Yuseong-gu, Daejeon 34055, Republic of Korea}

\author{Gungwon Kang}
\thanks{Corresponding author, gwkang@cau.ac.kr}
\affiliation{Physics Department, Chung-Ang University, Seoul 06974, Republic of Korea}

\date{\today}% It is always \today, today,

\begin{abstract}
We have numerically investigated close scattering processes of two black holes (BHs). Our careful analysis shows for the first time a non-merging ringdown gravitational wave coming from dynamical tidal deformations of individual BHs during their close encounter. The ringdown wave frequencies turn out to agree well with the quasi-normal ones of a single BH in perturbation theory, despite its distinctive physical context from the merging case. Our study shows a new type of gravitational waveform and opens up a new exploration of strong gravitational interactions using BH encounters.
\end{abstract}

%Use showkeys class option if keyword
\maketitle

%\tableofcontents

\label{sec:intro}
\mysection{Introduction} 

Lots of numerical studies on binary black hole (BH) coalescence have primarily focused on quasi-circular cases presumably because they are the main sources of gravitational wave (GW) observations~\cite{LIGOScientific:2018mvr,LIGOScientific:2020ibl,LIGOScientific:2021djp}. As the sensitivity of GW detectors increases and various future observation plans are proposed~\cite{Reitze:2019iox,Punturo_2010,amaroseoane2017laser}, however, a broader range of GW sources becomes of interest, resulting in active studies on eccentric binary black hole (BBH) systems in both numerical and approximation methods. In particular, highly eccentric BBHs and 
scattering BHs on hyperbolic orbits have been investigated~\cite{Gold:2012tk,Bae:2017crk,Bae:2020hla,Damour:2014afa,PhysRevD.94.104015,PhysRevD.96.064021,PhysRevD.97.044038,PhysRevD.103.104021,PhysRevD.106.024042,PhysRevD.107.064051,Cho:2018upo,Cho:2021onr,Cho:2022pqy}.

The numerical study on close encounters of two BHs is very interesting. The BBH merger can be regarded as an extreme case of this close encounter. Studying scattering systems, by suitably adjusting the impact parameters and speeds of the encountering BHs, can extend the scope of investigation beyond typical BBH merger scenarios. 
This broader examination may provide new insights into strong gravitational interactions between BHs. For instance, one may see how the horizons of individual BHs deform without reaching to a coalescence.

At the perturbative regime, it has been of interest whether or not a BH horizon could be tidally deformed~\cite{Binnington:2009bb,LeTiec:2020spy}. 
In the strong field regime, on the other hand, authors in Ref.~\cite{Prasad:2021dfr} have considered BH source's moments from numerical simulations of BBH mergers, indicating non-vanishing tidal Love numbers for BH horizons.
For a close scattering process of BHs, consequently, one naturally raises the possibility of dynamical tidal deformations of individual BHs involved. Moreover, it may lead to an additional generation of ringdown-like GWs from each scattered BH without coalescence. In this work, we observe that tidal deformations indeed occur during the close scattering of two BHs, producing ringdown GWs even if BHs do not merge. The leading frequencies of such ringdown GWs show remarkable agreements with those of quasi-normal mode (QNM) excitations for individual BHs in the perturbation theory.

\begin{figure}[ht]
\centering
\includegraphics[scale=0.27]{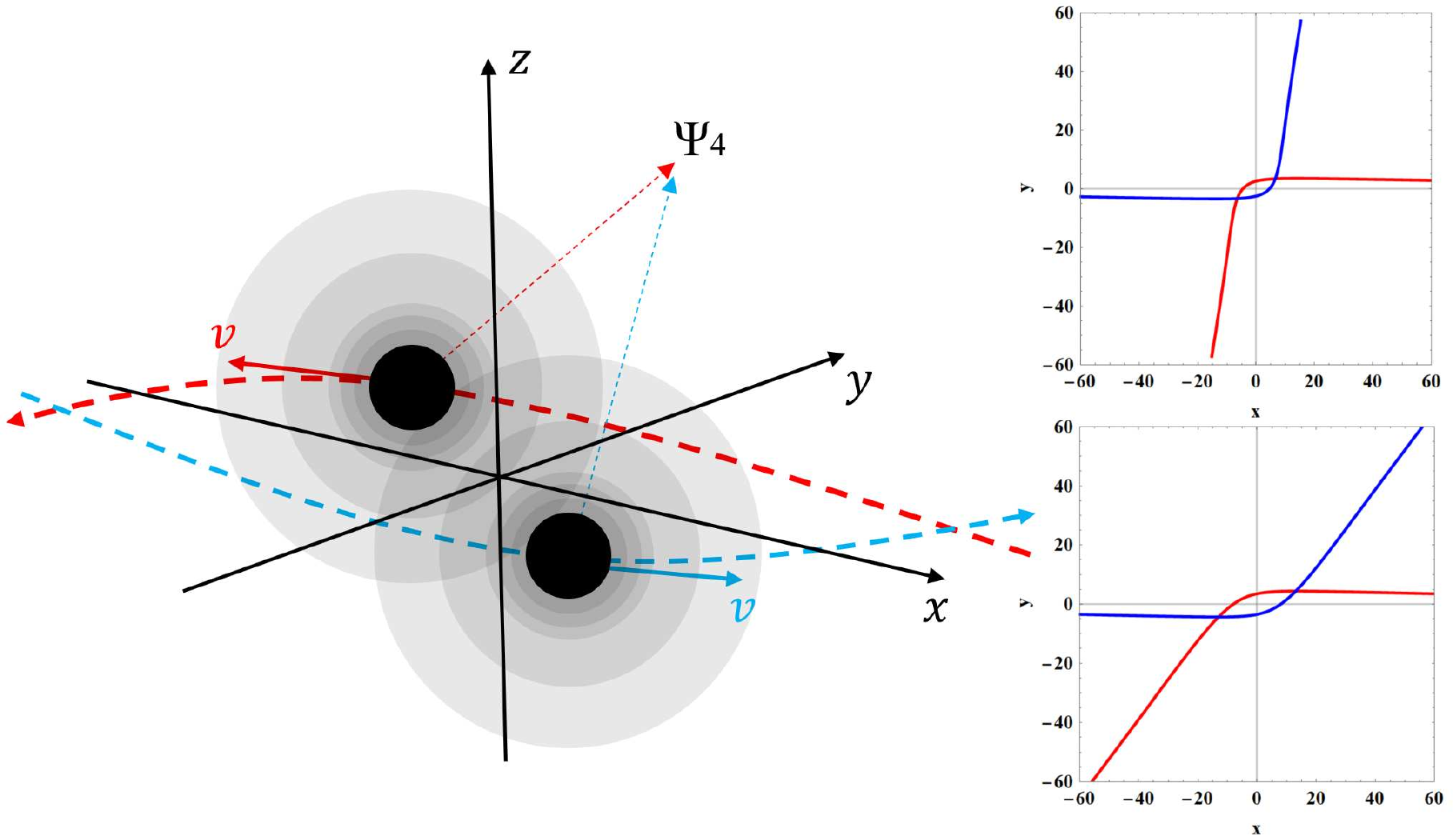}
\centering  
\caption{Two BHs in a hyperbolic encounter (left), and their trajectories on the $x$-$y$ plane for $b=8\,M$ (right top) and $b=10\,M$ (right bottom), respectively.}
 \label{fig:image2}
\end{figure}

\mysection{Numerical Simulation of 
Scattering Black Holes}

For the numerical relativity (NR) simulations, we use \textsc{EinsteinToolkit}~\cite{EinsteinToolkit:2023_05,Loffler:2011ay}, which is one of the most widely used open-source codes based on the Cactus framework. \textsc{McLachlan}~\cite{Brown:2008sb} is used for time evolution of the system with 8th order spatial finite difference method, and \textsc{Carpet}~\cite{Schnetter:2003rb} for the adaptive mesh refinement of the numerical grids. The computational domain covers up to 770 $M$ in geometrized units with radiative boundary conditions for an outer boundary, and seven mesh refinement levels are used with the finest grid spacings for different resolutions, $h\simeq0.0137\,M$ and $h\simeq0.0156\,M$, around the BHs. 

We use \textsc{TwoPuncture}~\cite{Ansorg:2004ds} to set the initial conditions for two equal-mass and non-spinning BHs in hyperbolic motion during the whole scattering process. To avoid overlap between the physical wave and junk radiation, the initial positions of the two BHs are set at $(\pm X,Y,Z)=(\pm200,0,0)\,M$. The initial momenta are calculated using $(p_{x},p_{y},p_{z}) = \pm|\vec{p}|(-\sqrt{1-(b/(2X))^{2}}, b/(2X), 0)$, where $\vec{p}$ represents the initial momentum. We have set the initial momentum as $|\vec{p}|=0.2886751\,M$ which corresponds to $v\approx0.5$, and considered two different impact parameters $b=8\,M$ and $b=10\,M$. In all cases, the ADM energies are nearly identical to $M_{{\textrm{ADM}}}\approx1.1563547\,M$, with only a very small difference on the order of $10^{-8}$. Therefore, we conducted a total of four simulation models, using two different impact parameters and two different grid resolutions for each.

The Weyl scalar $\Psi_{4}$ has been extracted up to $l=16$ mode in spin-weighted spherical harmonics, and we set the several extraction surfaces of $\Psi_{4}$ up to $r_{\rm ext}=400\,M$, which is far enough away from the outer boundary so that the artificial effects from it can be negligible. 
A schematic picture of the simulation and the trajectories of the BHs are shown in Fig.~\ref{fig:image2}.

\begin{figure}[ht]
\centering
  \includegraphics[scale=0.34]{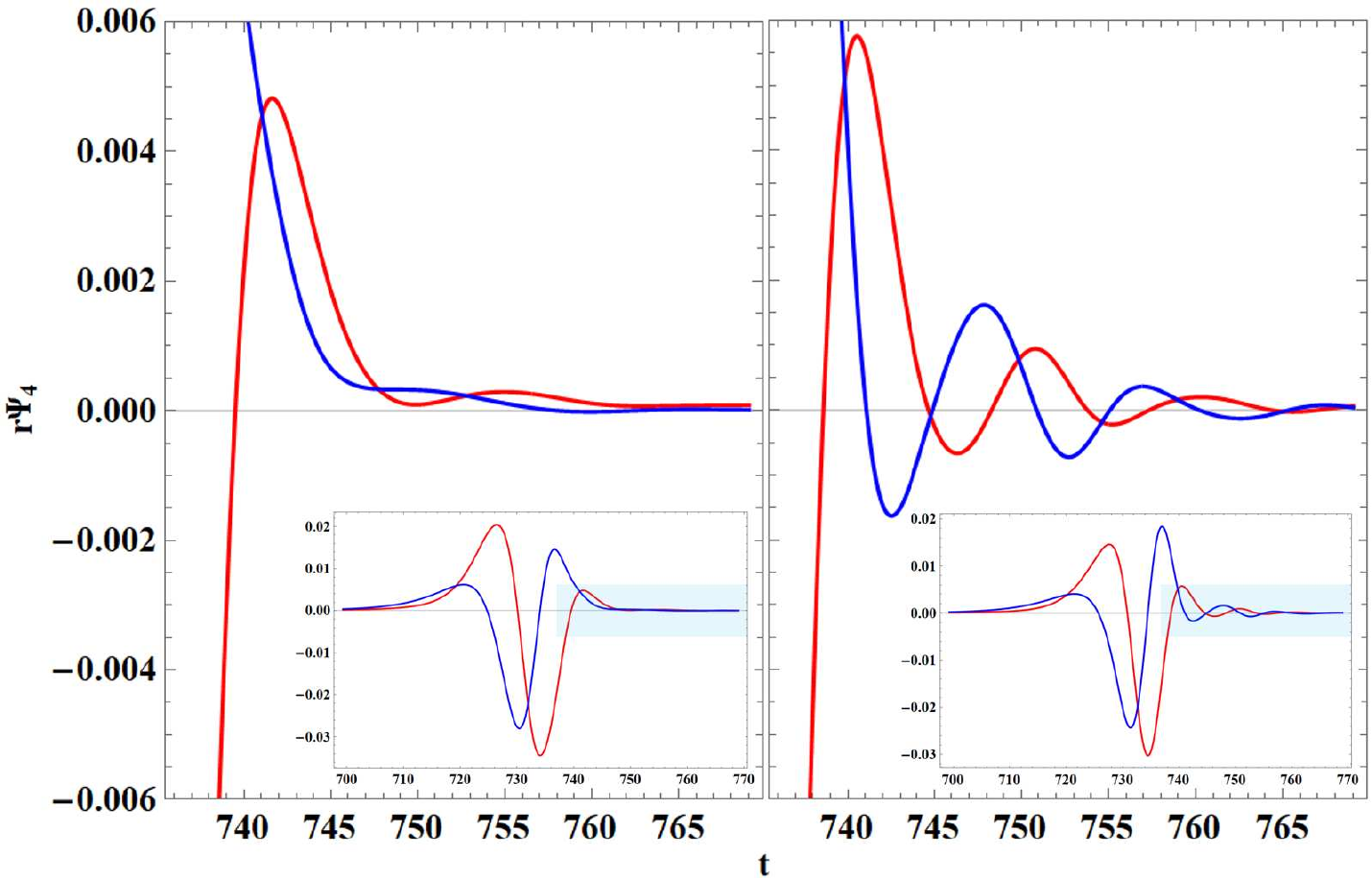}
   \caption{Waveforms in the outgoing phase after the close encounter (the parts shaded in blue in the inset figures) evaluated at $r_{{\textrm{ext}}}=300\,M$ for $b=8\,M$. Real (red) and Imaginary (blue) parts of $\Psi_{4}^{22}$ mode only (left) and the north pole observation $\Psi_{4}^{{\textrm{NP}}}$ of the sum up to $l=16$ (right), respectively.}
  \label{fig:22mode}
\end{figure}

\mysection{Identification of Ringdown Waves from Scattering Black Holes}

Unlike the case of a merging quasi-circular BBH, which is characterized by the IMR phases, it is convenient to divide the GW radiation from a scattering BH into ingoing and outgoing phases. Such division can be done by identifying either half of the scattering angle, the minimum relative distance of encountering BHs, or the maximum strength of radiated gravitational waveforms. In addition to the generation of GWs associated with its trajectory, tidally perturbed BHs may generate a ringdown-like gravitational radiation during a close encounter as speculated above.

The left panel of Fig.~\ref{fig:22mode} shows the radiated waveform of the $\Psi_{4}^{22}\equiv \Psi_{4}^{l=2,m=2}$ mode, which is believed to be most dominant typically in the merging case. One can see a weak, but oscillating signal in the outgoing phase, which is about $2 \, \%$ of the largest strength of the whole signal. In addition, we find that the multipole contributions higher than $l=2$ are not suppressed significantly in this case of close encounter. Actually, by summing up the numerical data obtained for several leading multipole modes, one can show that a much clear ringdown-like wave appears following the simple wave associated with the encountering orbit. In order to show that this ringdown-like wave is indeed originated from the dynamical tidal deformations of individual BHs, however, we carry out a more careful analysis as below.

Note first that, unlike the merging BH at rest, two BHs producing ringdown waves are moving fast. When observed at the extraction surface, the radiations from individual BHs could get blue/red-shifted, causing modulations in general. 
Secondly, ringdown waves coming from each BH will get deformed due to different positions and curvature effects such as gravitational lensing and absorption into BHs. Note, finally, that the positions of radiating BHs are not identical to the center of the finite extraction surface. Such center deviations, together with all effects mentioned above, could cause even a single $(l,m)$ mode radiation defined with respect to the center of each BH appear to be a sum of various modes on the extraction surface whose harmonics are defined with respect to the center of mass.

Instead of taking into account all aforementioned effects on multipole modes, we consider a GW as a whole at some special observing point minimizing these effects.
Let us consider the GW which is the summation of all modes up to $l=16$ as follows: 
\begin{align}
&\Psi_{4}(t,r,\theta,\varphi) \approx \sum_{l=2}^{16}\sum_{m=-l}^{l}\Psi_{4}^{lm}(t,r)_{-2}Y_{lm}(\theta,\varphi) \, ,
\end{align}
where ${}_{-2}Y_{lm}$ represents the spin-weighted spherical harmonics. When evaluating it at several positions on the extraction surface, one can see much larger and more clearly oscillating waves than that of the $(2,2)$ mode in Fig.~\ref{fig:22mode}. Note, however, that, due to the reasons mentioned above, most of these waves still possess some modulations, which makes the analysis in detail being difficult. 
We find that these modulations are minimized at the symmetric locations, {\it e.g.,} north/south poles. The Weyl scalar $\Psi_{4}^{{\textrm{NP}}}=\Psi_4(t,r,\theta=0,\varphi)$ at the north pole is given by the right panel in Fig.~\ref{fig:22mode}. Its strength reaches about $10 \, \%$ of the largest strength of the whole signal. In this way, we identify the ringdown wave coming from scattering BHs.

\begin{figure}[ht]
\centering
\includegraphics[scale=0.31]{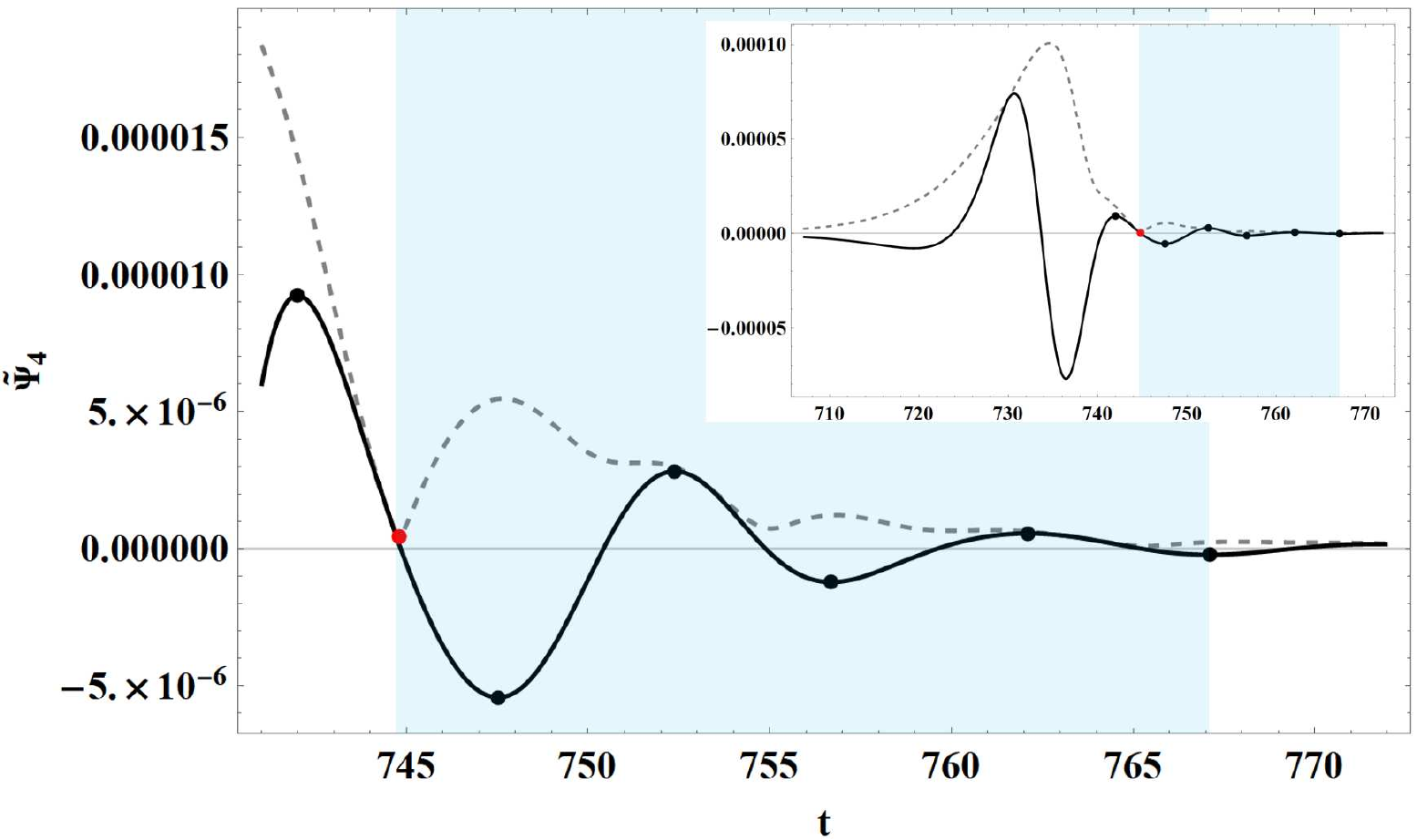}
\caption{Waveforms of $\widetilde{\Psi}_{4}$ (solid line) and 
$|\Psi_{4}^{{\textrm{NP}}}|$ (dashed line) for $b=8\,M$ at $r_{{\textrm{ext}}}=300\,M$ in the outgoing phase, respectively. 
Extremum points are denoted by bold dots. Ringdown waves used for QNM analysis is indicated by the sky-blue area.}
\label{fig::peaks}
\end{figure}

In Fig.~\ref{fig:22mode}, it is of interest to note that the magnitude of the third peak in Re[$\Psi_{4}^{{\textrm{NP}}}$] (colored red) is larger than that of the second one, which is unusual for a pure ringdown wave. This behavior turns out to be due to the gravitational radiation associated with the trajectory of two BHs, which leads to a decomposition of the signal given by
\begin{align}
&\Psi_{4}=\Psi_{4}^{{\textrm{TD}}}+\Psi_{4}^{{\textrm{RD}}},
\end{align}
where $\Psi_{4}^{{\textrm{TD}}}$ is the trajectory-driven (TD) wave, and $\Psi_{4}^{{\textrm{RD}}}$ the ringdown (RD) wave. In order to minimize $\Psi_{4}^{{\textrm{TD}}}$, we use a new coordinate system with a rotation where the new axis is aligned as close as the asymptotic outgoing trajectory which is a nearly straight line as can be seen in Fig.~\ref{fig:image2}.  
Consequently, at least one of the polarizations in $\Psi_{4}^{{\textrm{TD}}}$ rapidly decays to zero, enabling us to approximate the $\Psi_{4}^{\textrm{TD}}$  using a certain type of decaying function.
Fig.~\ref{fig::peaks} shows the corresponding polarization, $\widetilde{\Psi}_{4}\equiv {\textrm{Im}}[e^{2i\theta_{\textrm{out}}}\Psi_{4}^{{\textrm{NP}}}]$, in such a new frame with the rotation $\theta_{\textrm{out}}$.
This is the gravitational waveform to be used for further analysis of the RD wave below.

\mysection{Quasi-Normal Mode Analysis of Scattering Ringdown Wave}

When a single isolated BH is weakly perturbed, its gravitational radiation can be expressed as a sum of its characteristic excitations, the so-called QNMs. The RD wave contained in Fig.~\ref{fig::peaks} may be analyzed in terms of such characteristic modes. It should be pointed out, however, that the present case is very much different from the typical one in QNM analysis. Namely, a perturbed BH has a non-negligible equal mass BH nearby. Furthermore, two BHs are in motion with varying speeds, which might reveal very different characteristic excitations from the known ones for a single stationary perturbed BH.
Nevertheless, we will still use, for comparison, QNMs of a single isolated BH to approximately describe the ringdown wave in this complicated situation. It will be interesting to see how well such an approximation works.

Note that even a single QNM wave coming from the individual BH will get deformed when it is measured at the north pole due to various effects aforementioned such as boost, curvature and interference effects. The interference effect due to two identical BHs as different wave sources may be negligible because of the symmetrical measuring point. The curvature effect may also be neglected because of the clean waveform without small modulations as can be seen in Fig.~\ref{fig::peaks}. 
The boost effect will mainly come from the time dilation between the comoving and north pole observers. The effect due to the line-of-sight motion will be very weak because the coordinate distances of two BHs are within $\sim 15\,M$ from the center of mass, which are very small compared to those of the north pole observers. Furthermore, this effect become negligible when extrapolation to an observer at infinity is performed using different extraction surfaces ({\it i.e.,} $200\,M - 400\,M$). Notice that the speeds of fly-by BHs actually vary during the outgoing phase, resulting in a time-varying frequency at the north pole $\omega^{{\textrm{NP}}}(t) \simeq \omega_{{\textrm{QNM}}}/\gamma(t)$ with the QNM frequency measured by the comoving observer $\omega_{{\textrm{QNM}}}$. However, our estimation of the Lorentz factors $\gamma(t)$ based on numerical trajectories of two BHs shows that its maximum variance is within about $2 \%$. Thus, we assume a constant Lorentz factor, $\tilde{\gamma}$, for the late part of the encounter, namely, effectively regarding the fly-by BHs at a uniform motion within the time range of our analysis.
Hence, $\omega^{{\textrm{NP}}}\equiv\omega_{{\textrm{QNM}}}/\tilde{\gamma}$ is approximated.

Therefore, the whole ringdown part of the GW obtained at the north pole may be expressed as
\begin{align}\label{eq_Psi4ringdown}
&\widetilde{\Psi}_{4} =\sum_{i}  A_{i} e^{-\omega^{(i)}_{{\textrm{I}}}(t-\Delta t_{i})}\cos (\omega^{(i)}_{{\textrm{R}}}(t-\Delta t_{i}))\nonumber\\
&~~~~~~~+\frac{B}{(t-\Delta t_{{\textrm{T}}})^{p}}+C\,,
\end{align}
where the terms in the first line, $\widetilde{\Psi}_{4(i)}^{\textrm{RD}}$, represent radiation solely from the tidal deformations of the BHs. The damping part in the subsequent terms, $\widetilde{\Psi}_{4}^{\textrm{TD}}$, describing the TD waves is assumed to be a power-law decay. In this study, we carry out both single-mode ($i=1$) and double-mode ($i=1,2$) analyses. Our numerical results show that the spin parameters of the final BHs after encounter are negligibly small, {\it e.g.,} $\sim 10^{-4}$. Thus, we regard that the RD radiation comes from a tidal deformation of each non-rotating BH, {\it e.g.,} a Schwarzschild BH, in motion. Furthermore, it is assumed that dominant/sub-dominant radiations are due to the excitation of fundamental $l=2,3$ QNMs.

\begin{figure}[ht]
\centering
\includegraphics[scale=0.207]{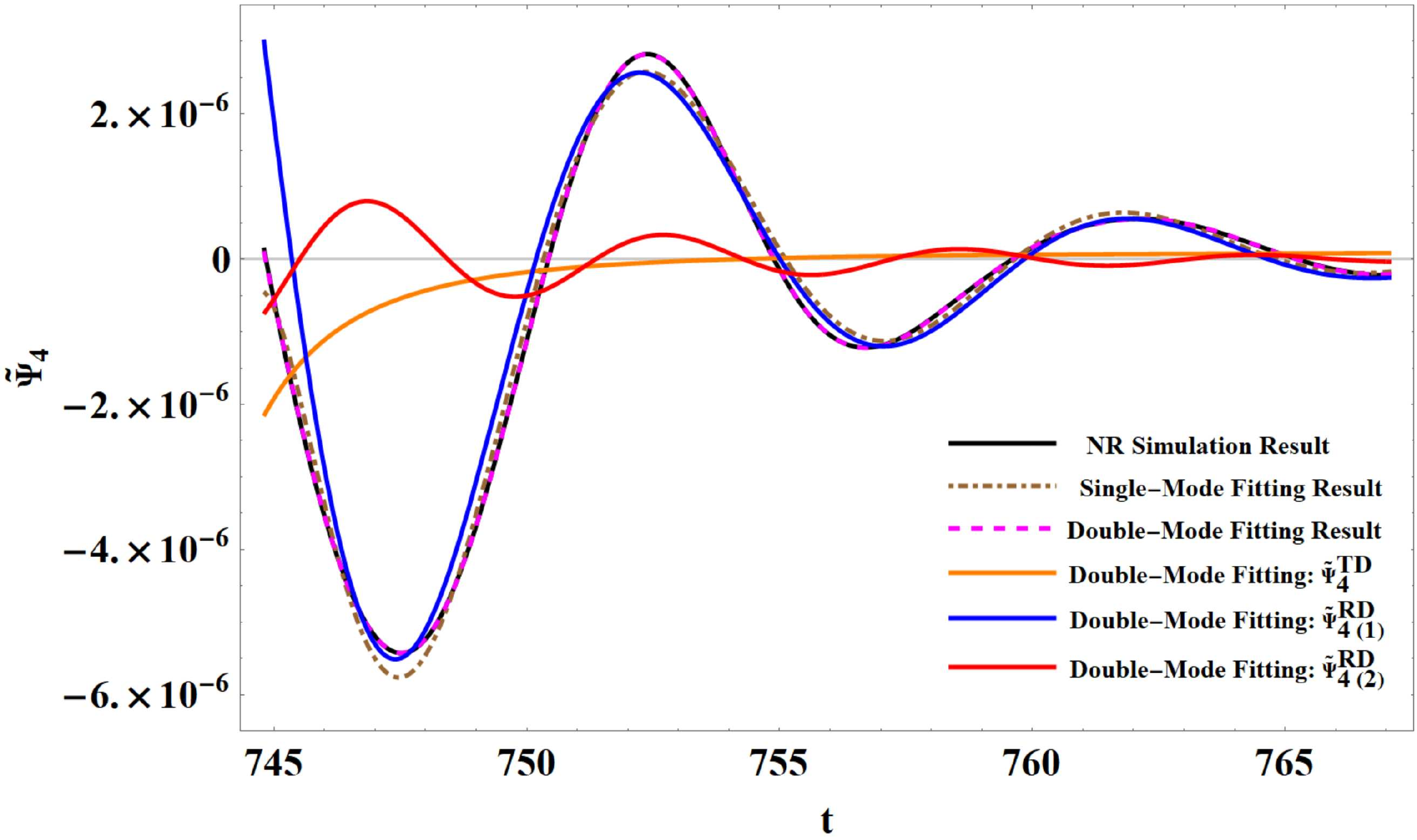}
~~\includegraphics[scale=0.207]{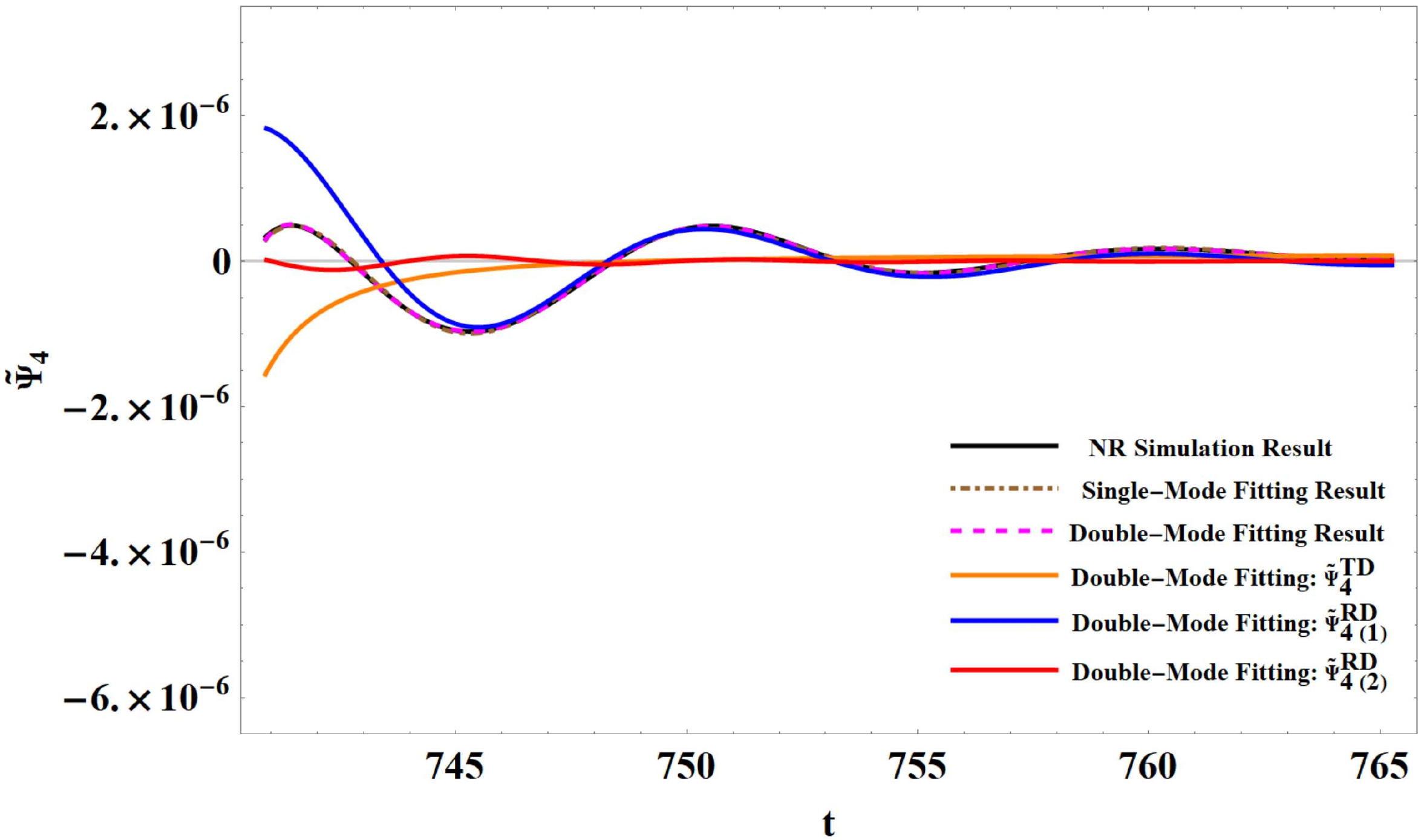}
\caption{Parameter fitting results for $\widetilde{\Psi}_{4}=\widetilde{\Psi}_{4}^{{\textrm{TD}}}+\widetilde{\Psi}_{4}^{{\textrm{RD}}}$ (dashed lines) at $r_{\textrm{ext}}=300\,M$ from the simulation with impact parameters of $b=8\,M$ (upper) and $b=10\,M$ (lower).}
\label{fig::fitting}
\end{figure}

To fit the numerical signal shown in Fig.~\ref{fig::peaks} with the formula in Eq.~(\ref{eq_Psi4ringdown}), we use the shaded time interval. The endpoint of this interval is chosen right before the numerical noise becomes comparable. The starting point was chosen where the RD signal becomes comparable to the TD contribution, which will be the local minimum of $|\Psi_{4}^{{\textrm{NP}}}|$ (dashed line) denoted by the red-dot in Fig.~\ref{fig::peaks}.
We performed a nonlinear regression using Newton's method, which iteratively refines the parameter estimates to minimize the sum of squared residuals.

\begin{table*}[ht]
\begin{center}
\begin{tabular}{c@{\hspace{0.5em}}c@{\hspace{1.5em}}c@{\hspace{1.5em}}c@{\hspace{1.5em}}c@{\hspace{1.5em}}c}
\toprule
\multicolumn{2}{c}{\textit{\textbf{RD Frequencies}}} & $\boldsymbol{M\omega^{(l=2)}_{{\textrm{\textbf{R}}}}}$ & $\boldsymbol{M\omega^{(l=2)}_{{\textrm{\textbf{I}}}}}$ & $\boldsymbol{M\omega^{(l=3)}_{{\textrm{\textbf{R}}}}}$ &  $\boldsymbol{M\omega^{(l=3)}_{{\textrm{\textbf{I}}}}}$ \\
\midrule
\multicolumn{2}{c}{\textit{Perturbation Theory}} & $0.3737$ & $0.0890$ & $0.5994$ & $0.0927$ \\
%\addlinespace
\midrule
\multirow{2}{*}{\textit{NR~$(b=8)$}~\,} & \textit{single-mode} & $0.3915(91)~(+4.8\,\%)$ & $0.0906(45) ~(+1.8\,\%)$ & -- & -- \\ 
& \textit{double-mode} & $0.3798(11)~(+1.6\,\%)$ & $0.0894(4)\,~~(+0.5\,\%)$ & $0.5965(234)~(-0.5\,\%)$ & $0.0617(234)~(-33.4\,\%)$ \\ 
\addlinespace
\multirow{2}{*}{\textit{NR~$(b=10)$}}  & \textit{single-mode} & $0.3738(14)~(+0.0\,\%)$ & $\,~0.0737(68) ~(-17.2\,\%)$ & -- & -- \\ 
& \textit{double-mode} & $0.3741(25)~(+0.1\,\%)$ & $0.0826(31)~(-7.2\,\%)$ & $0.6498(541)~(+8.4\,\%)$ & $0.0549(778)~(-40.7\,\%)$ \\ 
\bottomrule
\end{tabular}
\caption{Ringdown frequencies numerically obtained from scattering BHs. The percentages represent deviations from the theoretical values, $(M\omega)_{\textrm{QNM}}$, for a single Schwarzschild BH.}
\label{table:2}
\end{center}
\end{table*}

The fitting results for the cases of $b = 8\,M$ and $b = 10\,M$ at the extraction radius $r_{\mathrm{ext}} = 300\,M$ are illustrated in Fig.~\ref{fig::fitting}. As the impact parameter increases from $b=8 \, M$ to $b=10 \, M$, one can see that the strengths of both RD and TD contributions decrease, presumably, due to the less-strong interactions for the larger impact parameter. Notice that the weakening of the TD radiation is not severe relative to that of the RD one. Note also that the sub-dominant RD contributions are suppressed to about $13\,\%$ and $7\, \%$ of the dominant ones for the cases of $b=8\, M$ and $b=10\, M$, respectively. It can be seen in Fig.~\ref{fig::fitting} that the TD contribution still remains clearly even in the outgoing phase.

The RD frequencies obtained in this way depend on the coordinate systems used in numerical simulations in general. The gauge-free frequencies $\omega_{\infty}$ can be obtained at spatial infinity, which is done by extrapolation using a least squares method. This was achieved with a polynomial function of $1/r_{\textrm{ext}}$ for the extraction radii of $200\,M \le r_{\textrm{ext}}\le 400\,M$. The polynomial's order was chosen such that the extrapolated values matched up to the 10th significant digit.

Now, let us relate these frequencies $\omega_{\infty}$ to the dimensionless QNM ones $M_{{\textrm{BH}}}\omega_{{\textrm{QNM}}}$. Note that
\begin{align}
&M_{{\textrm{BH}}}\omega_{{\textrm{QNM}}}=(\tilde{\gamma} M_{{\textrm{BH}}})\bigg(\frac{\omega_{{\textrm{QNM}}}}{\tilde{\gamma}}\bigg) \equiv  M_{\textrm{boost}}\omega_{\infty}.
\end{align}
Here $M_{{\textrm{boost}}}\equiv \tilde{\gamma} M_{{\textrm{BH}}}$ is 
the boosted mass of the individual BH with respect to the north pole observer at infinity. This may be approximated as follows:
\begin{align}
&M_{\textrm{boost}}\simeq \gamma_{\infty}M_{BH} = (M_{{\textrm{ADM}}}-\Delta E)/2~.
\end{align}
Here $\gamma_{\infty}$ is the Lorentz factor of the individual BHs when they get separated away sufficiently, which differs from $\tilde{\gamma}$ about $1.5\,\%$. $\Delta E$ is the radiated energy through GWs in total. 
Our numerical results give $\Delta E=0.01312$, $0.00507$ for $b=8\,M$, $10\,M$, respectively. Using the Richardson extrapolation with a conservative assumption of 4th-order convergence for the results from the different resolutions, finally, we present the RD frequencies, $M_{\textrm{boost}}\omega_{\infty}$, in Table~\ref{table:2}. The numbers in parentheses represent the differences in the last digits from the high-resolution results, which can be regarded as finite difference errors.

In Table~\ref{table:2}, the comparison is given for these RD frequencies obtained from the NR waveforms and the QNM ones of a single Schwarzschild BH in perturbation theory~\cite{Berti:2005ys}. For the case of $b=8 \, M$, our single-mode fitting shows that the tidally-deformed RD wave could be due to the excitation of the fundamental $l=2$ quasi-normal mode within about $(4.8, 1.8) \, \%$ for the oscillatory and damping parts, respectively. Actually, the QNM frequencies of the overtone mode or higher multipole mode give much larger disagreement, excluding such modes as the dominant excitation. For the possibility of sub-dominant QNM excitations, our double-mode fitting shows that $l=2, 3$ fundamental modes are excited in agreement within about 
$(1.6, 0.5) \, \%$ and $(0.5, 33.4) \, \%$, respectively. Note that the agreement with the $l=2$ mode is improved. 

For the case of a less strong encounter with $b=10 \,M$, 
we observe slight enhancement in the agreements of real frequencies for the dominant mode. This enhancement might be because two BHs are in the linear regime more than the case of $b=8\,M$. Or, it could be simply because, as can be seen in Fig.~\ref{fig::fitting} (lower), the subdominant mode is too weak so that the dominant mode fitting is already good enough. Such behavior may also result in a larger deviation for the subdominant mode.
For the imaginary frequencies, on the other hand, the deviations become larger than those in the $b=8 \, M$ case. 
Note first that the agreement would depend on how accurately the TD wave can be subtracted from the whole ringdown signal. We speculate that the decaying behavior of the RD wave is more sensitive than its oscillatory one. Consequently, the agreements in imaginary frequencies would be worse than those in real frequencies as can be seen in Table~\ref{table:2}.
As mentioned above, the relative suppression of the TD wave is not severe in the case of $b=10\, M$, which likely leads to larger deviations in the imaginary frequencies.

Even though the physical situations for the RD radiation in a hyperbolic encounter are quite different from the merging case as mentioned above, it is interesting to see in the analysis above that the RD frequencies are in fairly good agreement with those QNM frequencies in the perturbation theory for a single static BH.

\mysection{Conclusions}

Our numerical study has shown for the first time that 
close hyperbolic encounters of two BHs could produce a non-merging ringdown gravitational wave coming from dynamical tidal deformations of individual BHs. The strength of such ringdown wave in $\Psi_4$ is about $10 \, \%$ compared to that of the trajectory-driven radiation. Our careful analysis shows that these waves can be regarded as characteristic excitation of the fundamental quasi-normal modes of each scattering BH.

The occurrence of such scattering ringdown waves is essentially due to the tidal deformations caused by companion BHs. The strength and characteristics of these tidal deformations, accordingly, ringdown radiations, can be controlled by adjusting the impact parameter and initial speeds of two BHs. Therefore, one may investigate much more various features of finely-tuned ringdowns than in the cases of binary BH coalescence. It also indicates that close encounters could provide another good way of testing strong gravitational interactions in general relativity. 

Our study shows a new type of gravitational waveform although extremely close and highly fast BH encounters would rarely occur in realistic astrophysical situations.
If observed, however, the ringdown wave could carry useful information about the scattering system.
As explained above, encountered BHs radiating ringdown waves are moving away, and so the effects of companion BH and their speeds can be encoded into the propagation of ringdown waves. Consequently, the detailed analyses of the radiations in terms of modes and directions could provide a good observational tool for distinguishing the source's system parameters such as its inclination angle and velocities of each BHs with respect to the center of mass.

Finally, it will be of interest to see how this non-merging ringdown radiation behaves in more general cases of BH scatterings with unequal mass, non-vanishing spins and their various alignments, a wider range of impact parameters, and initial speeds. Similarly, it will also be interesting to study non-merging ringdown radiations from close encounters of two compact objects including a neutron star, white dwarf, and BH. Our study opens up a new exploration of strong gravitational interactions using BH encounters.

\vspace{10pt}

\begin{acknowledgments}
The authors thank Chan Park, Jinho Kim, Sang Hoon Oh and Hyung Mok Lee for their valuable discussions.
Y.-B.B. was supported by IBS under Project Code No. IBS-R018-D1. Y.-H.H. and G.K. are supported in part by the National Research Foundation of Korea (NRF) funded by the Ministry of Education (NRF-2021R1I1A2050775) and (NRF-2022R1I1A207366012), respectively. 
Y.-B.B. and G. K. are supported in part by the National Research Foundation of Korea(NRF) grants funded by the Ministry of Science and ICT (No. NRF-2021R1F1A1051269) and (NRF-2021R1A2C201247313), respectively.
Computing resources and technical support are provided by the IBS Research Solution Center, KISTI Supercomputing Center (No. KSC-2020-CRE-0352), and KASI’s gmunu HPC clusters.
\end{acknowledgments}

%\appendix
%\nocite{*}
\bibliographystyle{apsrev4-1}
\bibliography{ref}% Produces the bibliography via BibTeX.

\end{document}